\documentclass[%
 reprint,
 superscriptaddress,
 amsmath,amssymb,showpacs,
 aps,
 prb,
]{revtex4-1}

\usepackage{graphicx}
\usepackage{dcolumn}
\usepackage{bm}
\usepackage{mathrsfs}
\usepackage{float}
\usepackage{color}

\begin{document}

\preprint{APS/123-QED}

\title{Line defects in Graphene: How doping affects the 
electronic and mechanical 
properties}

\author{Daniel Berger}
 \email{daniel.berger@math.ucla.edu}
 \affiliation{Department of Mathematics, University of California, Los Angeles, CA 90095, United States}
\author{Christian Ratsch}
 \affiliation{Department of Mathematics, University of California, Los Angeles, CA 90095, United States}
 \affiliation{Institute for Pure and Applied Mathematics, University of California, Los Angeles, CA 90095, United States}

\date{\today}

\begin{abstract}
Graphene and carbon nanotubes have extraordinary mechanical and electronic properties. Intrinsic line defects such as local non-hexagonal reconstructions or grain boundaries, however, significantly 
reduce the tensile strength, but feature exciting electronic properties. Here, we address the properties of line defects in graphene from first-principles on 
the level of full-potential density-functional theory, and assess doping as one strategy to strengthen such materials. We carefully disentangle the global and 
local effect of doping by comparing results from the virtual crystal approximation with those from local substitution of chemical species, in order 
to gain a detailed understanding of the breaking and stabilization mechanisms. We find that n-type doping or local substitution with nitrogen
increases the ultimate tensile strength significantly. In particular, it can stabilize the defects beyond the ultimate tensile strength of the pristine 
material. We therefore propose this as a key strategy to strengthen graphenic materials. Furthermore, we find that doping and/or applying external stress 
lead to tunable and technologically interesting metal/semi-conductor transitions.

\end{abstract}

\maketitle

\section{Introduction}

Graphene-based materials have attracted wide interest not only for its exquisite electronic properties\cite{novoselov2005two,zhang2005experimental,RevModPhys.81.109}, but also for the extraordinary tensile strength
of graphene and carbon nanotubes \cite{Lee18072008,stankovich2006graphene}.
Although much progress has been made towards production of pristine graphene sheets by chemical vapor deposition~\cite{Li05062009,doi:10.1021/nl801827v,doi:10.1021/nl101629g}, grain-boundaries and line defects are 
still unavoidable for graphene sheets with sizes interesting for applications.~\cite{huang2011grains} Such one-dimensional defects form when crystal growth starts from multiple centers, as the crystal structures 
are very unlikely to have the exact same orientation. Because of its low-dimensionality and the rigid bonding structure of graphene the structural variety along these lines are
typically governed by the presence of heptagon-pentagon (h-p) and octagon-pentagon (o-p) defects.~\cite{doi:10.1021/nl100988r}
Grain boundaries have been studied extensively and are predicted to have distinct electronic~\cite{PhysRevB.79.195429,yazyev2010electronic,huang2011grains}, magnetic ~\cite{vcervenka2009room}, 
and chemical properties~\cite{PhysRevB.81.165447}. But they are also known to reduce the overall tensile strength of the overall graphene sheet.~\cite{wei2012nature,PhysRevB.81.195420} 

A recent study by Bisset \text{et al.}\cite{doi:10.1021/nn404746h} has shown that the catalytic reactivity of defected graphene can be increased by an order of magnitude when applying mechanical strain. 
Maximizing the ultimate strength of defected graphene is necessary to optimize it towards a versatile energy material.
A fundamental understanding of the electronic and mechanical properties of such defects is therefore essential. 


Line defects, continuous h-p and o-p reconstructions, have only recently gained interest as alternative and controllable quasi one-dimensional defects in graphene.\cite{lahiri2010extended,PhysRevLett.106.136806,doi:10.1021/acsnano.5b01762,C4CP02552K} 
These line defects can be specifically engineered from the lattice mismatch of graphene with the substrate during 
vapor deposition.~\cite{lahiri2010extended} Recent STM measurements have already shown that the the o-p line defect has remarkable electronic properties~\cite{lahiri2010extended,PhysRevLett.106.136806}. However, its
mechanical properties remain largely unknown.~\cite{yazyev2014polycrystalline} 

Altering the electronic and corresponding mechanical properties through substitutional doping has been proposed as one strategy to tailor graphene devices towards specific functionalities.
Substitutional doping with nitrogen has been realized in the lab through chemical vapor deposition or electrochemical treatment.~\cite{doi:10.1021/nl803279t, Wang08052009,doi:10.1021/cs200652y}
Its effect on the electronic properties of pristine or point-defected graphene has been intensively studied from both the experimental and computational side.~\cite{rani2013designing,
doi:10.1021/nn9003428, PhysRevB.84.245446,PhysRevB.86.045448}. However, relatively few studies have addressed doping in extended line defects~\cite{PhysRevB.85.035404,C4CP02552K}. First-principle calculations of 
the substitution formation energies by Brito \textit{et al.}\cite{PhysRevB.85.035404} have shown that nitrogen doping is thermodynamically more favored in the vicinity of a line defect compared to the
graphene region. This suggests that actually doping at such line defects might be more relevant than in the pristine region itself, which motivates a detailed study of doping effects at such line defects.



In this work we address the effects of nitrogen and boron doping on the electronic and mechanical properties of h-p and o-p line defects in graphene by means of full-potential density-functional theory. 
This technique is free of empirical parameters and has been proven reliable in the prediction of mechanical properties~\cite{Carter08082008}, especially when an accurate description of 
the chemical bonding conditions up to the level of bond breaking is necessary. 
We scrutinize the undoped pristine and line-defected structures and investigate the effect of external strain and then study the effect of doping on the tensile strength and electronic structure of line-defected 
graphene.  We study the effect of \textit{local} substitutional as well as conventional \textit{global} doping: local doping is studied in terms of substitutional doping with nitrogen and boron at all distinguished pentagon, heptagon and
octagon sites. Conventional global doping, the change of concentration of quasi-free charge carriers, a.k.a. a shift of the Fermi level, is simulated in the \textit{virtual crystal approximation}~\cite{vegard,SCHEFFLER1987176},
in which carbon atoms are replaced by virtual chemical species with fractional nuclear charges and fractional electron count. 
We find that both the substitutional nitrogen doping as well as a global n-doping strengthens the material beyond the ultimate strength of the graphene structure through essentially the same mechanism, namely
localization of additional electrons within the defect region. We furthermore suggest doping as a key strategy for tailoring the rich electronic properties of line defects.


%

\section{Computational Method}

All electronic structure calculations are done on the level of spin-unrestricted density-functional theory within the full-potential, all-electron framework of \texttt{FHI-aims}~\cite{blum2009ab}. 
Electronic exchange and correlation is treated on a semi-local level using the PBE functional~\cite{PBE1996}.
Default \textit{tight} integration settings and the \texttt{FHI-aims} \textit{tier2} basis set ensure well converged total energies.
Reciprocal space is sampled very accurately such that the number of Monkhorst-Pack k-grid~\cite{PhysRevB.13.5188} points along each unit cell axis times the length of the lattice vector is
always larger than 40\,{\AA}.
In order to fully decouple periodically reproduced sheets a vacuum of 40\,{\AA} has been used throughout.
All atoms have been fully relaxed until residual forces are smaller than 10$^{-2}$\,eV/{\AA}. These computational setting reproduce the experimentally observed C-C distance of 1.42\,{\AA} for pristine graphene.
Dispersion interaction is known to be crucial for an accurate description of interactions perpendicular to the graphene plane.~\cite{gobre2013scaling} All in-plan properties are, however, 
dominated by the strong covalent C-C bonds. Inclusion of ab-initio dispersion correction~\cite{Tkatchenko2009} showed indeed no effect on the geometry and structural stability of graphene and is therefore neglected in this study.

\begin{figure}
\centering
\includegraphics[width=6.5 cm]{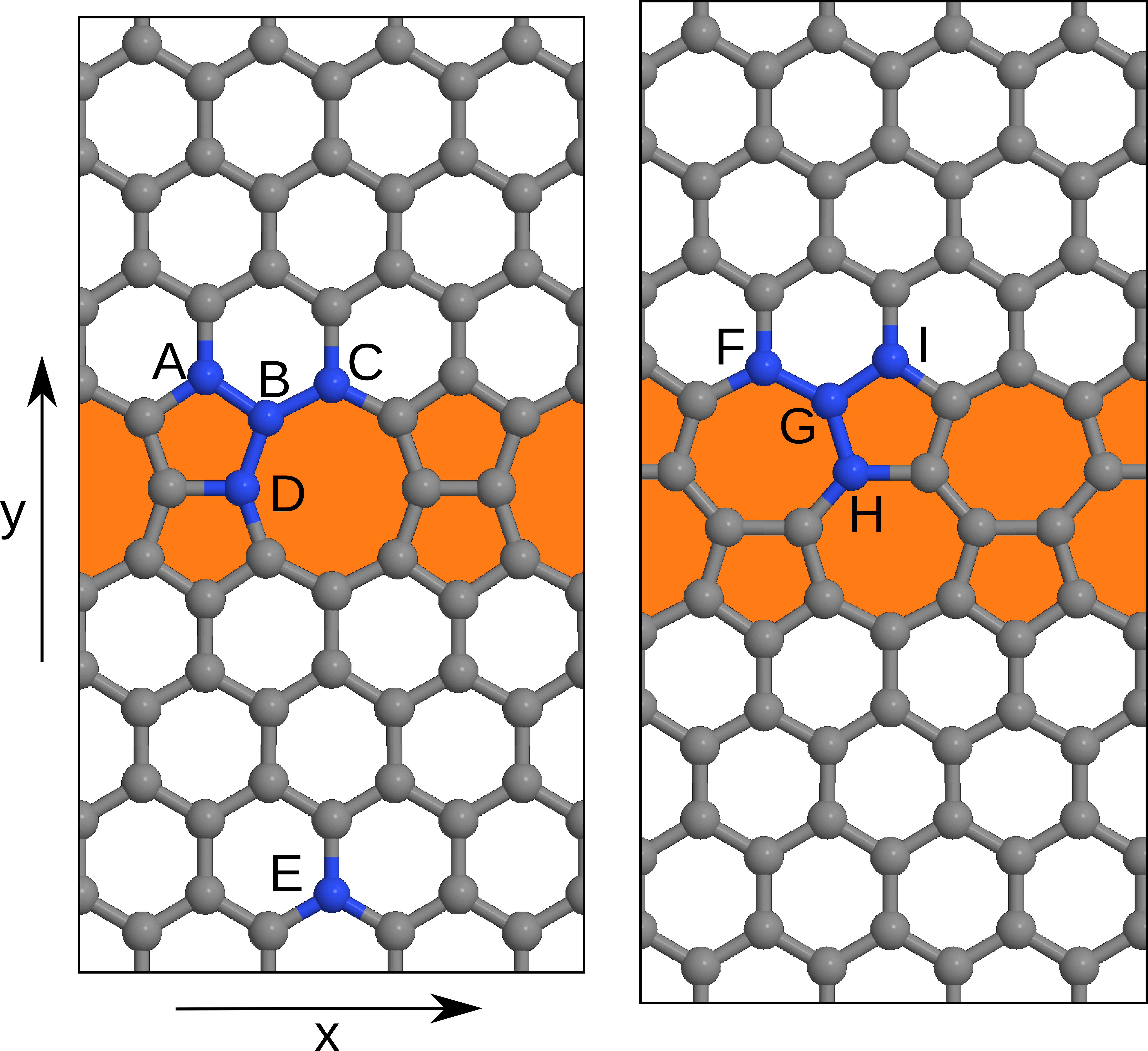}
\caption{Employed supercells of the octagon-pentagon (left) and heptagon-pentagon (right) line defect structures. Doping sites are labeled and highlighted in blue.}
\label{fig1}
\end{figure}

Employed supercells of the defected graphene structures have been increased systematically in size to minimize finite size effects such as the structural relaxation around the 
defect. Band structures of the line defects were calculated in a supercell irreducible in the direction of the line defect (one octagon and two pentagons, respectively two heptagons and two pentagons), while line 
defects are separated by 13 hexagons.
Calculations of the substitution energetics and all mechanical properties of the octagon-pentagon (heptagon-pentagon) defect were performed in a supercell with two octagons and four pentagons (four heptagons and four 
pentagons) and seven rows of hexagons
separating periodically reproduced line defects, as shown in Fig.~\ref{fig1}. These supercells that comprise 68 (72) atoms are necessary to yield converged substitution energies within 10 meV, which has been confirmed with even larger supercells 
with 9 rows of hexagons. Reference calculations of the pristine graphene were performed in a well converged 64 atoms supercell.

The strain-stress relation is determined by stretching the unit-cell perpendicular to the line defect while relaxing all internal degrees of freedom.
The stress-tensor was converged until all entries were less than 10$^{-6}$\,eV/{\AA}$^{3}$, throughout.
The mechanical properties are studied in terms of the Young's modulus $\gamma$ and the ultimate strength (or breaking strength) $\sigma_{max}$.
The Young's modulus is calculated as the second derivative of the total energy $E_{tot}$ with respect to the external strain $\epsilon$
evaluated at zero strain:
\begin{equation}\label{eq2}
 \gamma=\frac{1}{V_0} \frac{\partial^2 E}{\partial \epsilon^2 }\bigg|_{\epsilon=0}\;\;\;\;.
\end{equation}
With $V_0$ being the volume of the material (at zero strain), $\gamma$ is rigorously only defined in three dimensions.
Assigning it to a 2D material involves one free parameter, namely the effective thickness of the sheet $d_{eff}$. 
We use the experimental interlayer distance in graphite for $d_{eff}$, 3.35\,{\AA} \cite{Lee18072008}.
Equivalently, $\gamma$ can be calculated directly from the stress-tensor $\sigma$ computed with \texttt{FHI-aims} as
\begin{equation}
 \gamma=\frac{z}{d_{eff}} \frac{\partial \sigma}{\partial \epsilon}\bigg|_{\epsilon=0}\;\;\;\;,
\end{equation}
with z being the dimension of the supercell perpendicular to the sheet. This method yields essentially the same results.

The free energy balance for the substitution of carbon atoms with nitrogen and boron are approximated on the level of self-consistent total energies. The substitution energy~\cite{PhysRevB.38.7649} per atom
\begin{equation}\label{eq1}
 E_f^i(X)=\frac{1}{n}\Big(E^i(X) - E_{undoped} -n\cdot \mu_X +  n\cdot \mu_C\Big)\;\;\;\;,
\end{equation}
is defined as the total energy of the system with substitutional species $X$ at site $i$, $E^i(X)$, referenced against the total energy of the undoped system, $E_{undoped}$, together with the
chemical potentials for removing a carbon atom $\mu_C$ and adding an atom of species $X$, $\mu_X$. Analogous to Brito \text{et al.}\cite{PhysRevB.85.035404}, $\mu_N$ is taken as half of the energy of the nitrogen molecule
(closed-shell singlet), $\mu_B$ is the energy of a bulk atom in the alpha-boron conformation and $\mu_C$
is the energy of a carbon atom in the pristine graphene sheet. $n$ is the number of substituted equivalent sites.


\section{Results \& Discussion}

\subsection{Undoped graphene}\label{secIIIA}

\subsubsection{Electronic Properties}

\begin{figure}
\centering
\includegraphics[width=8.0 cm]{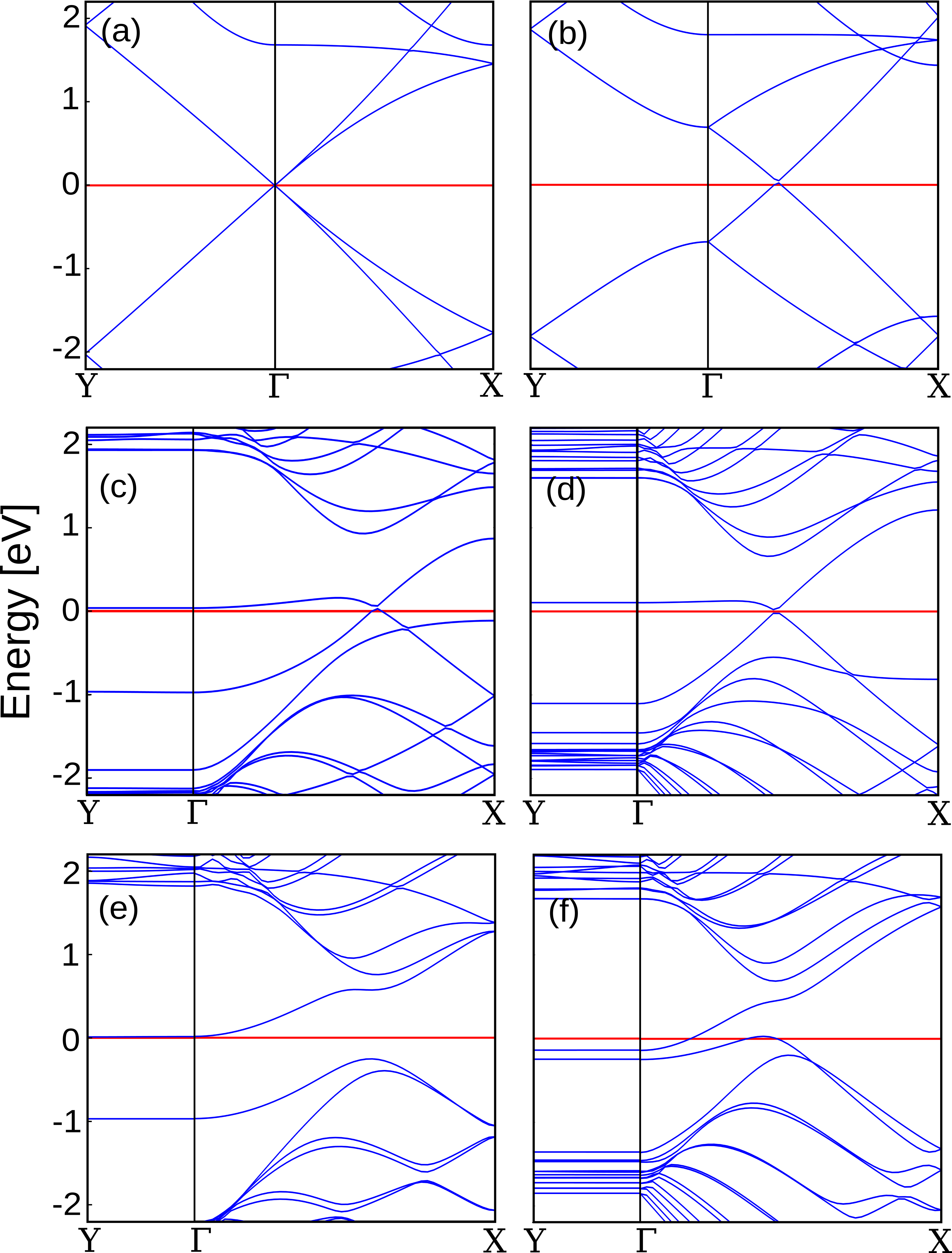}
  \caption{Electronic band structure of the undoped (a \& b) pristine graphene, (c \& d) octagon-pentagon line defect and (e \& f) heptagon-pentagon line defect, projected on the reciprocal supercell vectors,
  without external strain (a \& c \& e) and with 10~\% external strain along the y-direction (b \& d \& f). Reciprocal space was sampled very accurately on a k-grid mesh of 20$\times$20$\times$1, 40$\times$4$\times$1 and 40$\times$4$\times$1.
  Points X and Y refer to coordinates (0.5/0/0) and (0/0.5/0) in the coordinate system reciprocal to that defined in Fig.~\ref{fig1}. 
  The energy scale is referenced against the Fermi level (red line).} 
\label{fig2}
\end{figure}

Pristine graphene consists of sp$^2$ hybridized carbon atoms arranged in a honeycomb structure~\cite{RevModPhys.81.109}. Its electronic band structure exhibits states with a linear dispersion (Dirac cone). In our choice of a rectangular supercell
these states intersect the Fermi level in the center of the reciprocal space ($\Gamma$). Application of an external stress breaks the symmetry, which is reflected in a shift of the Dirac point in reciprocal space and a general
narrowing of the band structure (see Fig.~\ref{fig2}(b)).

The electronic band structures of line-defected graphene are qualitative different. The supercell approach periodically reproduces the line defect, which results in essentially flat bands perpendicular to the line 
defect ($\Gamma \rightarrow$ Y) for both types of line defects (Figs.~\ref{fig2}(c)-(f)). Only in the direction of the line defect ($\Gamma \rightarrow$ X) the true features become evident.
The band structure of the o-p line defect displays two important characteristics, namely two flat band regions close to the $\Gamma$
point, which lead to a remarkable high density of states close to the Fermi level, which makes it interesting for many technological applications by itself. The fact that two bands are crossing the Fermi energy
furthermore make it a potential 1-dimensional electric wire, as discovered by Lahiri \textit{et al.}~\cite{lahiri2010extended}.
Our spin-unrestricted calculations showed negligible spin density in the whole system, giving no reason to expect any ferromagnetic effects which have been predicted for the line defect in graphene 
nanoribbons~\cite{PhysRevB.84.075461}.

As pointed out by Lahiri \textit{et al.}~\cite{lahiri2010extended} the flat band region is a true feature of the defect and does not arise from zigzag edge states, which would appear between 2$\pi$/3 and $\pi$ along 
the $\Gamma \rightarrow$ X axis \cite{PhysRevB.54.17954}. Our Mulliken charge analysis\cite{Mulliken1955} shows that the flat band region close to the $\Gamma$ point arises from a localized state in a narrow 
region around the line defect, predominantly on sites A,B \& C (shown in Fig.~\ref{fig1}), which is in very good agreement with STM measurements~\cite{lahiri2010extended,:/content/aip/journal/apl/101/11/10.1063/1.4752441} and
recent first-principle simulations~\cite{C3RA41815D}.
The band region closer to the $X$-point arises almost entirely from contributions of sites D, and is therefore equivalent to the zigzag edge state.
Upon external strain the o-p line defect maintains its overall features (Fig.~\ref{fig2}(d)), however, with a slight shift of the Fermi-energy and a narrowing of the gap between valence and conduction band.
Strain-induced spin-polarization as reported for the line-defect in graphene nanoribbons~\cite{Qu2015116} has not been observed.

The h-p line defect, on the contrary, exhibits a band gap in the direction of the line defect, as shown in Fig.~\ref{fig2}(e). The band touching the Fermi-level at the $\Gamma$-point is sharply localized on sites H and 
neighboring zigzag edge sites. As can be seen in Fig.~\ref{fig2}(f), this band crosses the Fermi level when an external stress is applied 
leading to weak electric conductivity along the line defect. We therefore predict that the electric properties of the h-p line can be mechanically switched between conducting and semi-conducting behaviour, suggesting interesting electronic applications.

A Mulliken charge analysis displays only minor charge redistributions within the defect region. Sites B,C \& D (H \& F) each attract 0.01 electrons from sites A (I). 
These values are in excellent agreement with recent first-principles calculations by Ren \textit{et al.}~\cite{PhysRevB.91.045425}.
The defect region (all labeled sites) is overall charged with 0.025 electrons. The counter charge is homogeneously distributed over all carbon atoms in the pristine region suggesting a small but 
long range effect through the presence of the line defect.
We find similar charge distributions for the h-p line defect, with electrons accumulating on the atoms H and F.
In both cases, external strain increases the amount of the charge redistribution. The overall charge within the defect region, however, stays the same.

\subsubsection{Mechanical Properties}

The well known exceptional mechanical properties of graphene are characterized through a large Young's modulus
and large ultimate strength. 
We calculate the Young's modulus for pristine graphene as 1.00\,{TPa}, both for the armchair and the zigzag direction. This is in excellent agreement with experiment (1.0 $\pm$ 0.1\,{TPa}) \cite{Lee18072008} and preceding ab-initio 
simulations (1.05\,{TPa}) \cite{PhysRevB.81.195420} and molecular dynamics simulations employing empirical bond order potentials (1.01 $\pm$ 0.03\,{TPa}) \cite{doi:10.1021/nl901448z}.

\begin{figure}
\centering
\includegraphics[width=8.5 cm]{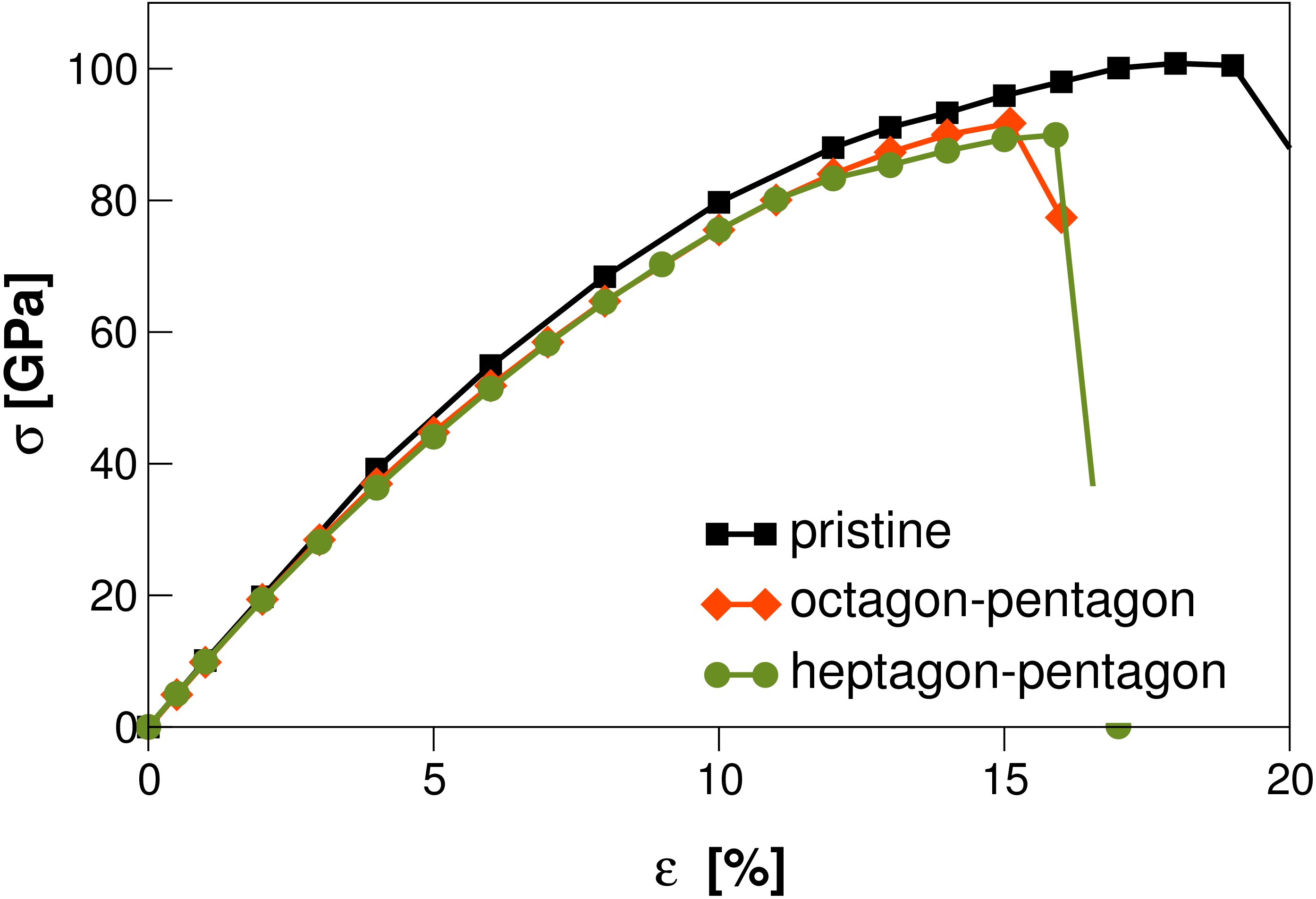}
\caption{Calculated tensile stress as a function of the applied strain $\epsilon$ for the octagon-pentagon and heptagon-pentagon line-defected structures as shown in Fig.~\ref{fig1},
compared to the (armchair) pristine graphene reference.} 
\label{fig3}
\end{figure}

The ultimate strength $\sigma_{max}$, the maximum of the stress-strain curve in Fig.~\ref{fig3}, is calculated as 101\,{GPa} at an external strain of 0.183 for the armchair direction (112\,{GPa} at 0.234 external 
strain for zigzag, not shown in Fig.~\ref{fig3}). 
These values are again in very good agreement with experiment~\cite{wei2012nature} (100\,{GPa} and 118\,{GPa}, respectively), showing that the PBE functional allows for an accurate description of the
mechanical properties even far away from the equilibrium structure. Differences in the mechanical properties between zigzag and armchair can be explained with the different number of bonds along the break line.
With PBE lattice parameters the ratio of bonds per unit length of zigzag to armchair direction is 1.15 which very well matches the respective ratio of the ultimate strengths. 

We now focus on the two fundamental configurations, the h-p and the o-p line defect at the zigzag edge (Fig.~\ref{fig1}) in order to address the mechanical properties of the line-defected graphene.
The resulting stress-strain curves are shown in Fig.~\ref{fig3}. The Young's modulus is 1.01\,{TPa} for both kind of line defects, and is essentially equal to that of 
pristine graphene. Line defects therefore do not effect the elasticity of graphene, at least in the linear regime.
The maximum of the strain-stress curve in Fig.~\ref{fig3} defines the ultimate strength of the material. Further increasing the external strain induces a breaking of chemical bonds and hence ultimate degradation of 
the graphene sheet. The ultimate strength is calculated as 91.7\,{GPa} (89.9\,{GPa}) for the o-p (h-p) line defect, and is roughly 10\% below the pristine zigzag graphene reference. This shows that line 
defects are indeed weakening the overall strength of the material. 
We want to point out that the strain value at which the line defect breaks depends on the dimensions of the supercell, i.e. the number of hexagons separating periodic line defects, and does not reflect 
insufficient convergence.
Both line defects show slightly different breaking mechanisms. The h-p structure has its weak point at the bond between sites ``H'' connecting two pentagons (see Fig.~\ref{fig1}), where
all the stress is loaded onto this bond. On the contrary, the o-p structure has two similarly weak bonds connecting the two hexagon domains (bonds between equivalent sites B and D). 
This allows to balance the load between both bonds leading to a slightly higher ultimate strength. 



\subsection{Substitutional doping}
The advantage of a computational study of substitutional doping is that one can easily test every element in the periodic table.
Species whose covalent radii differ much from that of carbon are, however, 
unlikely to yield stable and planar structures. In fact, we tested substitutional doping with oxygen, phosphorous and silicon which all lead to very unstable structures. We, therefore, focus on nitrogen and boron as
two species with very similar covalent radii. Compared to carbon both species show opposite trends in ionization potential and electron affinities and should hence result in opposite doping conditions (n- and p-type).
Both line defects provide several distinct sites shown in Fig.~\ref{fig1} which were tested for substitutional doping at three different doping concentrations: 
single-atom substitution (n$_{min}$=1), substitution of half of all equivalent sites (1/2 n$_{max}$) and substitution of all equivalent sites 
(four-fold substitution for sites A,C,D,F \& I and eight-fold for site B,G \& H in the structures shown in Fig.~\ref{fig1}). Several different configurations are possible for 1/2 n$_{max}$ concentration. 
In the following we always report results for the lowest energy configuration. Except for boron doping of site D, maximized distances between doping atoms are always the energetically preferred configuration.

\begin{table*}
\setlength{\tabcolsep}{7pt}
\begin{tabular}{ll|| d d d d |d c|dddd}
\hline
$E_f$ [eV] & & A & B & C & D & E & $E^{*}$ & F & G & H & I \\ \hline
\hline
N & n=1               & -0.41  & -0.14  & 0.49  &  -0.41 &  0.56 & 0.89 & 0.68 & 0.03  & 0.42  & -0.15      \\ 
 & n=1/2 n$_{max}$     & -0.43  & -0.05  &  0.56 & -0.43  &   &  & 0.81 & 0.25 & 0.68  &  -0.17    \\ 
 & n=n$_{max}$      & -0.14  &  1.05  & 0.45  &  0.19  &   &   &  0.85  & 0.78  &  1.16  &  0.06     \\ \cline{2-12} 
  & Ref.~\onlinecite{PhysRevB.85.035404}  & -0.67  & -0.50  & 0.19  &  -0.84 &  0.41 & 0.63 &  &   &   &       \\ \hline \hline
B & n=1               & 0.83  & 0.11  & 0.10  & 0.27  & 0.73 &  0.75  & 0.24   & 0.46 & -0.17  & 1.15       \\
 & n=1/2 n$_{max}$     & 0.95  & 0.39  & 0.24 & 0.20  &   &  & 0.28 & 0.51  & -0.08  &  1.21     \\ 
 & n=n$_{max}$   & 0.91  & 0.82  & 0.35  & 0.48  &   &   &  0.45  & 0.89  & 0.65  &  1.34     \\ \cline{2-12} 
  & Ref.~\onlinecite{PhysRevB.85.035404}  & 1.02  & 0.21  & 0.36  &  0.47 &  0.98 & 1.0 &  &   &   &       \\ \hline \hline
\end{tabular}
\caption{Calculated substitution energies per atom (in eV) with nitrogen and boron at several distinct sites labeled in Fig.~\ref{fig1} and for different concentrations as defined in Eq.~\ref{eq1}. E refers to the substitution in the hexagonal area in the defected system, while E$^{*}$ 
denotes the substitution in pristine graphene.} 
\label{table1}
\end{table*} 
 
\subsubsection{Substitution energies}
\label{subsection:energies}

Boron and even more so nitrogen substitution is much more favored in the defect region than in the pristine region. This becomes evident in the single-substitution energies summarized in Table~\ref{table1}. 
Even in the pristine-like region in the defected system (site E) substitution is significantly more favored than the substitution in a pristine graphene supercell (labeled $E^{*}$ in Table~\ref{table1}). 
This reflects the long-range perturbation of nitrogen doping on the electronic structure discussed previously by Lambin \textit{et al.}\cite{PhysRevB.86.045448} and the long-range effects on the electronic 
structure induced by defects in graphene~\cite{PhysRevLett.105.196102}. We hence predict a gradient in the doping concentration towards the defect, which in turn leads to a macroscopic gradient in the electrochemical
potential.


Table~\ref{table1} shows that boron substitution is energetically less favored than nitrogen substitution. In fact, only site H is found to have a negative energy for single boron substitution. In the o-p line 
defect sites B and C show the lowest cost for boron substitution of $\sim$0.1 eV.
Nitrogen substitution is not only generally more favored for single substitution, but also allows for higher doping concentrations. In fact, nitrogen substitution at half the equivalent sites is favorable for sites 
A,B,D \& I, and site A even has a negative substitution energy in the high-concentration limit. The clear preference of site A for high doping concentrations can be explained with the large distance between
equivalent sites, minimizing the repulsive dopend-dopend interaction. 
Comparing the substitution energetics between both types of line defects shows that the h-p structure is in general less attractive to doping atoms. 
Here, only site I yields negative substitution energies for up to 1/2 n$_{max}$, and only a small positive value for the high concentration limit.
We observe the general trend that the substitution energy per atom increases with the concentrations, showing that dopend-dopend interaction for nitrogen and boron is indeed
repulsive. This is in agreement with previous studies of pristine graphene.~\cite{doi:10.1021/nn9003428,PhysRevLett.98.196803} Only in the case of boron doping at site D we see that substitution for mid 
concentrations becomes lower in energy when both boron atoms sit at the same octagon. 

Structural reorganization upon doping is very small, in general. It is $\sim$0.1\,{\AA} for boron and less than 0.04\,{\AA} for nitrogen substitution. Fully optimized structures reveal an average N-C (B-C) bond 
distance of 1.43\,{\AA}(1.50\,{\AA}) for the o-p line defect, and 1.42\,{\AA}(1.49\,{\AA}) for the h-p line defect. This trend is in very good agreement with other results in the literature~\cite{rani2013designing,PhysRevB.85.035404} 
and consistent with the covalent radii of N and B relative to the sp$^2$ hybridized carbon atoms~\cite{B801115J}. All equilibrium geometries remain planar except for the high boron concentration limit at sites H, 
where an undulative distortion occurs.

Our substitution energies reproduce the same relative stabilities of the sites as reported by Brito \textit{et al.} (Ref.~\onlinecite{PhysRevB.85.035404}). However, they
are systematically higher by 0.2-0.4\,{eV} for 
nitrogen and lower by 0.1-0.2\,{eV} for boron substitution. The systematic shift for the line defects and for the pristine graphene agree within 0.1\,{eV}, which points to a different
calibration of the chemical potentials. We emphasize that we use very tight convergence settings for the forces, k-grid and supercell sizes, and argue that the FHI-aims \textit{tier2} basis sets used in this study 
guarantee a very accurate description of the total energy, without applying any pseudopotential approximation for the core electrons. 
For the lack of previous studies, we cannot reference the h-p defect against literature values.

In the following study of the electronic and mechanical properties we focus on the lowest energy configurations (for a given concentration) as the thermodynamically most relevant candidates: site A \& D for
single substitution of nitrogen and nitrogen in the medium concentration limit 1/2 n$_{max}$, site A for the high concentration limit in the o-p line defect, and site I for nitrogen doping in the h-p line defect. Boron substitution is mostly unfavored 
energetically, but can be made thermodynamically feasible at high temperatures and/or adequate partial pressures. We therefore also discuss boron substitution for all concentrations and study sites B \& C for single substitution, 
sites C \& D for medium concentrations 1/2 n$_{max}$, site C for maximal concentration, and site H for the h-p line defect.

\subsubsection{Electronic structure}

\begin{figure}
\centering
\includegraphics[width=7.5 cm]{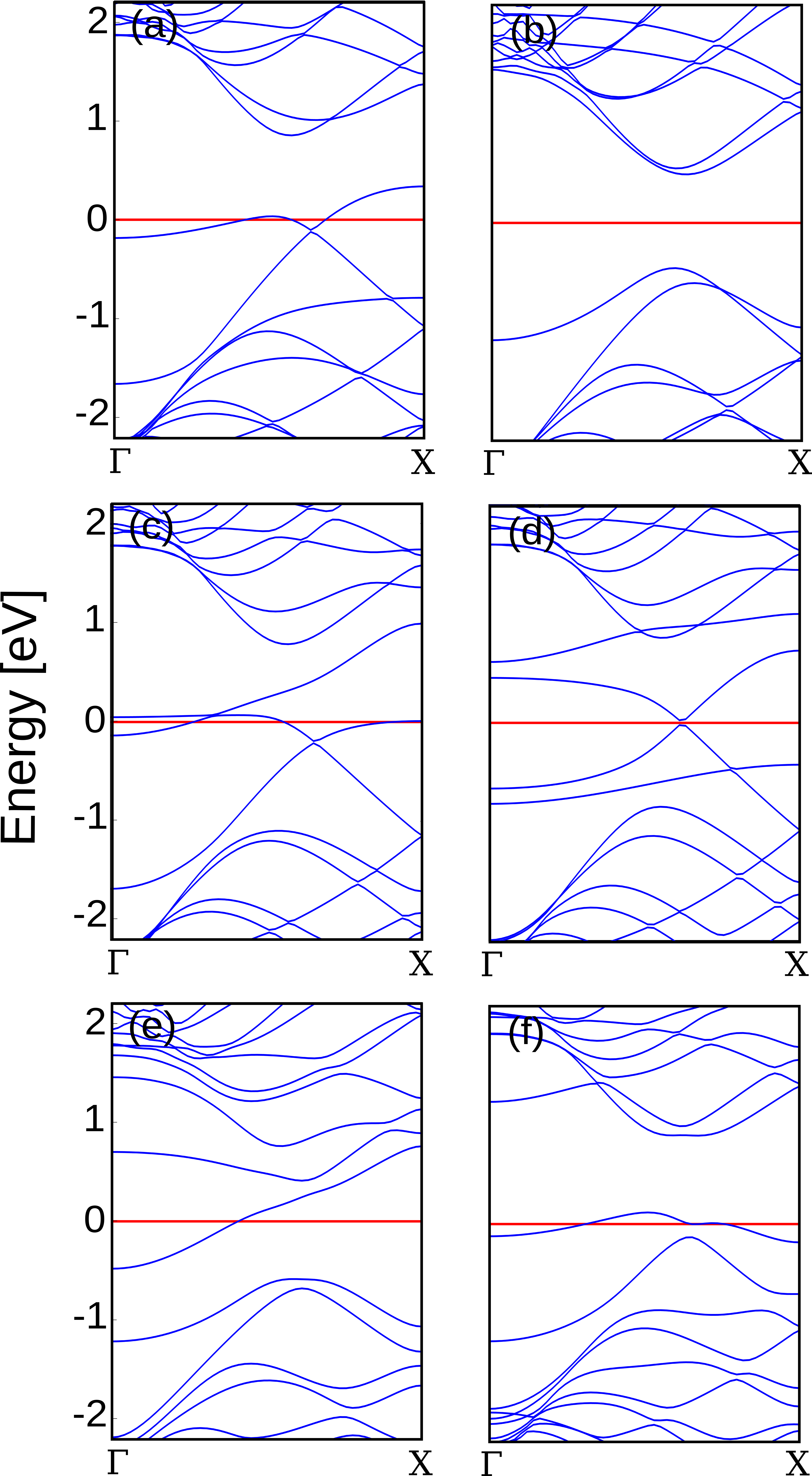}
  \caption{Electronic band structure of the (a-d) octagon-pentagon and (e-f) heptagon-pentagon line defects with (a \& e) low nitrogen and (c \& f) low boron concentration. Figures (b) and (d) show the octagon-pentagon 
  defect with high nitrogen and high boron concentrations. Reciprocal space was sampled on a k-grid mesh of 40$\times$4$\times$1 and 40$\times$4$\times$1. The band energies are referenced against the Fermi level (red line).} 
\label{fig4}
\end{figure}

The electronic structure for substitutional doped line defects is summarized in Fig.~\ref{fig4}. Low nitrogen concentrations in the o-p (Figs.~\ref{fig4}(a)) and h-p (Figs.~\ref{fig4}(e))
line defects essentially reproduce the electronic band structure of the undoped systems (Fig.~\ref{fig3}(c) \& (e)). The Fermi energy is shifted towards higher energy, reflecting n-type doped conditions
as expected from nitrogen doping in graphene \cite{doi:10.1021/nl803279t,doi:10.1021/ja402555n,doi:10.1021/nn1002425,PhysRevB.85.035404}. On the other hand, boron doping leads to significant changes in the band structure 
(Figs.~\ref{fig4}(c) \& (f)); e.g. additional bands crossing the Fermi level. 
This behaviour can also be seen by the charge transfer process with the substituted atom. While nitrogen only loses 0.01-0.02 electrons, charge transfer associated to boron is an order of magnitude larger, with an 
uptake of 0.2-0.4 electrons per boron atom. This suggests that boron atoms act as electron sinks, which substantially effects the mechanical properties discussed in Section \ref{subsec:doped_mechanics}.
This charge transfer reaction between doping atoms and line defects is long range, which becomes evident when studying nitrogen doping deep in the hexagonal domain (7 hexagons away from the defect).
In this case the charge transfer between the substitutional atom and its environment differs as well by an 
order of magnitude (-0.03 electrons for nitrogen, and +0.37 for boron). Strikingly, the effect on the charges in the defect region is a magnitude larger for nitrogen (0.12 more electrons) compared with boron (0.01 less
electrons). This suggests that the line-defects act as strong sinks for electrons (n-doping) but not for holes (p-doping).

With increasing concentration nitrogen doping leads to drastic changes in the electronic structure. Fig.~\ref{fig4}(b) shows the bands structure for the high concentration limit of substitution at site D. 
The former metallic o-p line defect now becomes semiconducting along the line defect. This transition is accompanied by a charge transfer of 0.05 electrons from the nitrogen atoms onto sites C.
Furthermore, the flat band regions characteristic for the o-p line defect completely disappear.
A corresponding metal/semiconductor transition has also been reported for pristine graphene when substituted nitrogen atoms form a line of dimers~\cite{:/content/aip/journal/apl/99/1/10.1063/1.3609243}.
The case of substitution at sites D in the high concentration limit essentially resembles such a line of nitrogen dimers. We therefore assign the observed metal/semiconductor transition to the line of such 
nitrogen dimers and not to the line defect. Nonetheless we would like to point out that the line of nitrogen dimers is a much more realistic scenario in the line defect than in pristine graphene,
since it yields much lower substitution energies.

\subsubsection{Mechanical properties}\label{subsec:doped_mechanics}

\begin{table*}
\begin{tabular}{ll|| d d |d}
\hline
$\sigma_{max}$ [GPa] & & A & D  & I  \\ \hline
\hline
N & n=1               & 91.2 & 88.2  &    90.3    \\ 
 & n=1/2 n$_{max}$     & 92.4  & 90.3  & 90.7       \\ 
 & n=n$_{max}$      & 96.7   & 100.7  &    93.9    \\  \hline
\end{tabular}
\vspace{0.5cm}

\begin{tabular}{ll|| d d d |dd}
\hline
$\sigma_{max}$ [GPa] & &  B & C & D & F & H \\ \hline
\hline
B & n=1               &  81.6 &  87.1 &   &   & 85.6         \\ 
 & n=1/2 n$_{max}$     &  & 84.9  &  84.1 &  &  80.8         \\ 
 & n=n$_{max}$   &  & 84.9  &   & 86.9  &       \\  \hline
\end{tabular}
\caption{Calculated ultimate strength of the doped octagon-pentagon and heptagon-pentagon line defect for the lowest substitutional energy conformation. The ultimate strength of the reference undoped system 
is 91.7\,{GPa} for the octagon-pentagon line defect, respectively 89.9\,{GPa} for the heptagon-pentagon line defect.} 
\label{table2}
\end{table*}

The effect of nitrogen and boron substitutional doping on the Young's modulus is negligible in both cases. Even for large concentrations the Young's modulus is calculated as 1.01 $\pm$ 0.02\,{TPa}.
In contrast to that, the overall strength of the material is significantly influenced by the species and concentration of the substitutional dopend. Table~\ref{table2} summarizes the ultimate strength of the doped 
line defects calculated for different doping concentrations in their lowest energy conformations (see Section~\ref{subsection:energies}). Nitrogen doping consistently leads to a higher ultimate strength than boron doping. 
In fact, boron doping does not promote the ultimate strength of the line defects at all, but rather decreases the strength significantly below the reference value of the undoped system. 
Contrary to that nitrogen doping stabilizes the material beyond the ultimate strength of the undoped system. We even observe an astonishing high ultimate strength of 101\,{GPa} for the substitution of site D in the 
high concentration limit. This is essentially equal to the ultimate strength of pristine graphene (100\,{GPa}, note that the direction of strain is to be compared to the armchair direction). 
Analysis of the individual bonding strengths reveals that in this case the bonds between the line defect and neighboring hexagons become the weakest, and not the defect structure itself. 

Our results highlight site A and D (I and H) as outstanding sites for substitutional nitrogen doping as they are energetically most favored and also promote the ultimate strength. In fact, we verified that doping at 
the energetically less preferred sites (e.g B and C) decreases the ultimate strength.
As discussed above, these two sites show the largest atomic charges in the undoped system without strain, and also observe the largest change in charge upon strain. 
Furthermore, structural relaxations upon nitrogen doping are almost negligible. This suggests that substitutional doping strengthens the material not through changing the character of chemical bonds but simply
through an increase of electron density in the defect region.
As the defect region acts as an electron sink, the latter could be also done through global (and not local substitutional) doping, i.e the global increase of quasi-free charge carriers, which 
may then accumulate in the defect region.
We will put this hypothesis to scrutiny in the following section.

\subsection{Doping with virtual species}

Within the limits of thermodynamical stability quasi-free additional electrons can be realized through homogeneous doping in the lab. Alternatively, such conditions can be generated through an application of an 
external voltage. An advantage of a computational study is that one is not constrained to what is chemically or experimentally feasible. One can even substitute with \textit{virtual} species with fractional nuclear 
charge and equivalent fractional number of electrons. This approach is especially straightforward to realize in a full-potential DFT framework like \texttt{FHI-aims} as no uncertainty can arise from scaling any 
pseudopotentials.

In order to reveal whether a local or global effect of doping leads to an enhanced ultimate strength, we perform two numerical experiments in which we dope with such \textit{virtual} chemical species.
In the first, we only substitute all equivalent sites of site A and D in the o-p line defect, respectively site H and I in the h-p line defect, with virtual chemical species between nuclear charges 5 and 7.4 and 
measure the resulting ultimate strength.\footnote{FHI-aims employs species specific numerical atomic orbitals, which are \textit{per se} not transferable across the periodic table. The \textit{tier2} basis set of carbon however allows a very accurate description
of nitrogen and boron, and \textit{vice versa}. Nonetheless, we checked the convergence of the total energy with respect to the basis set of virtual species by adding more basis functions,
i.e. to describe a virtual species with a nuclear charge
of 6.5 we specifically added nitrogen basis functions.} Z$_X$ will denote the proton and electron number of the virtual species at site $X$.
In a second experiment we simultaneously substitute all atoms in the system including all atoms in the hexagonal domain region. 
This procedure is known as the \textit{virtual crystal approximation}~\cite{vegard,SCHEFFLER1987176}.
Because of the high-concentration character of the doping in both experiments we can employ smaller supercells with half the size. 


\begin{figure}
\centering
\includegraphics[width=8 cm]{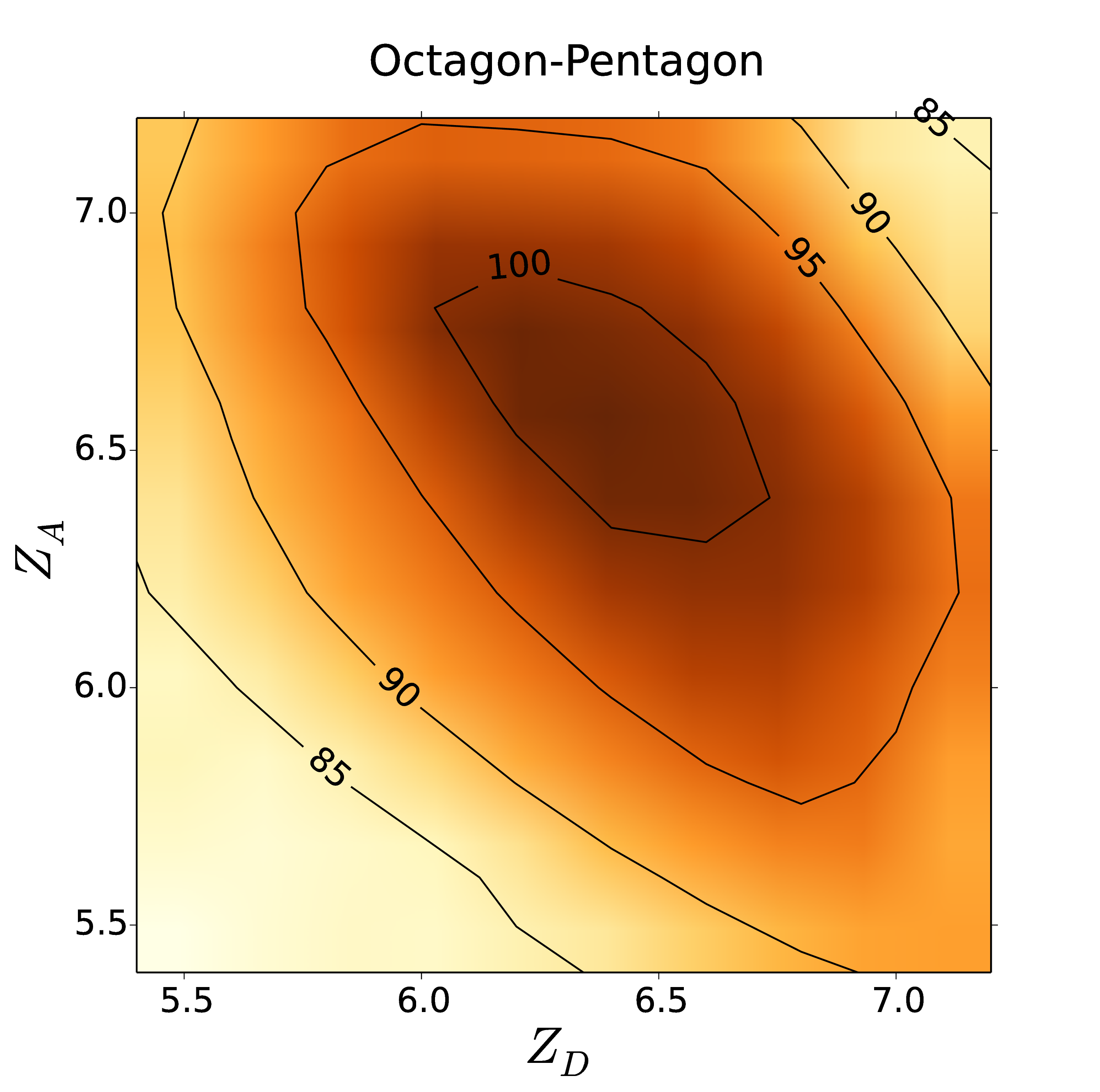}
\includegraphics[width=8 cm]{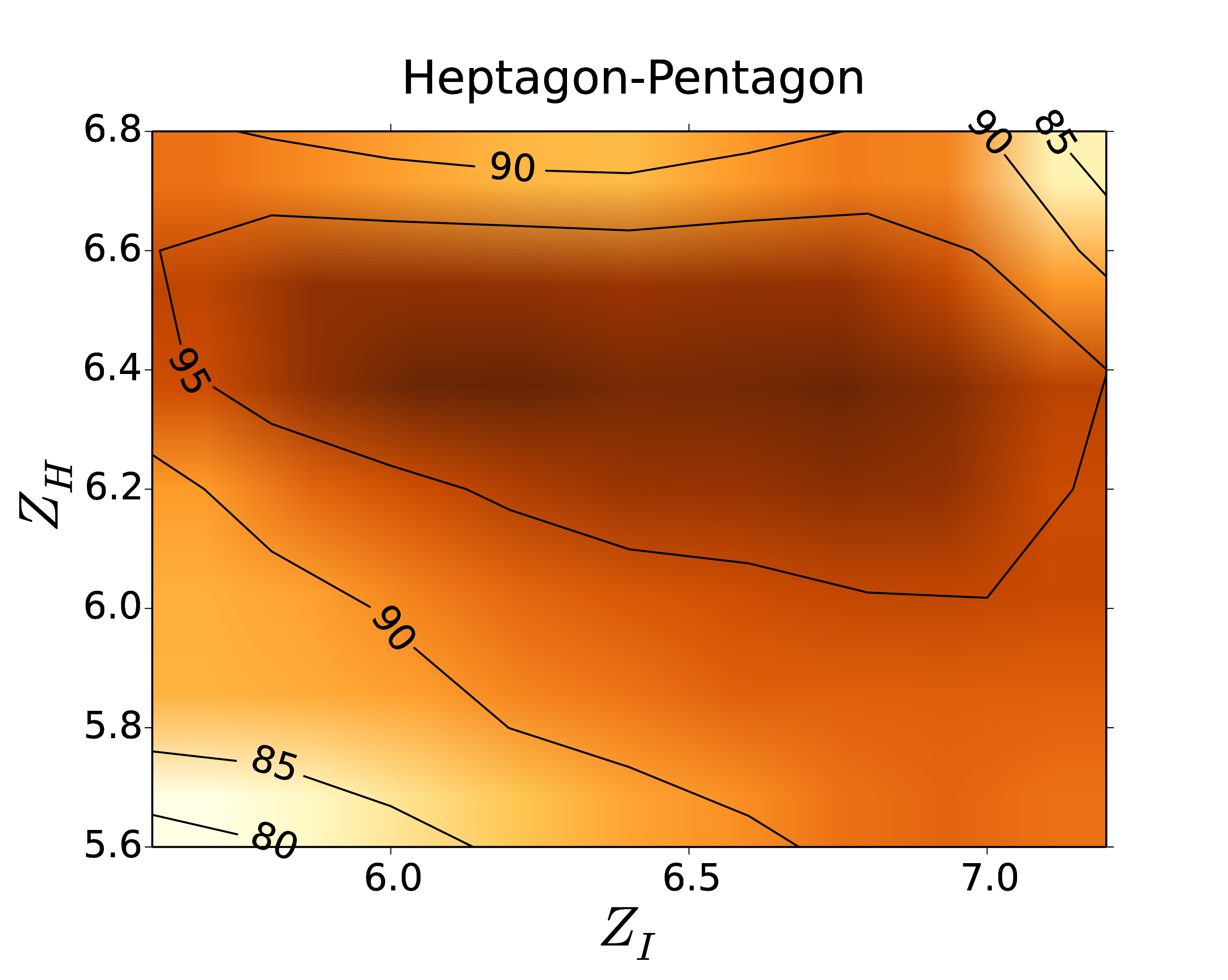}
  \caption{Contour plot showing the ultimate strength of the octagon-pentagon (heptagon-pentagon) line defect doped with virtual species at sites A and D (H and I) in GPa. Z$_X$ denotes the proton and electron
  number of the virtual species at site $X$. Calculation were performed in the high concentration limit of substitutional doping. A Gaussian smearing of 0.3 was used to smoothen the contours.} 
\label{fig5}
\end{figure}

Local substitution with virtual species reveals a broad peak around Z$_A$=Z$_D$=6.5 (Z$_H$=Z$_I$=6.5), whereas virtual species with $Z<$6 and $Z>7.2$ reduce the ultimate strength of the line defect, 
as shown in Fig.~\ref{fig5}. With 102\,{GPa} (100\,{GPa}) for the o-p (h-p) line defect the peak values lie above those for nitrogen substitution and for pristine graphene. 
Especially in the o-p case, where sites A and D are equal in numbers, the contour lines in Fig.~\ref{fig5}(a) are symmetric to the diagonal, i.e. symmetric with respect to exchange of A and D.
While site A and D have very different local environments, substitution at these sites has essentially the same effect. This suggests that the improvement of the mechanical properties is not caused by changes in
the local bonding structure, but simply through additional electrons which occupy the same orbital in the defect region.



\begin{figure}
\centering
\includegraphics[width=8 cm]{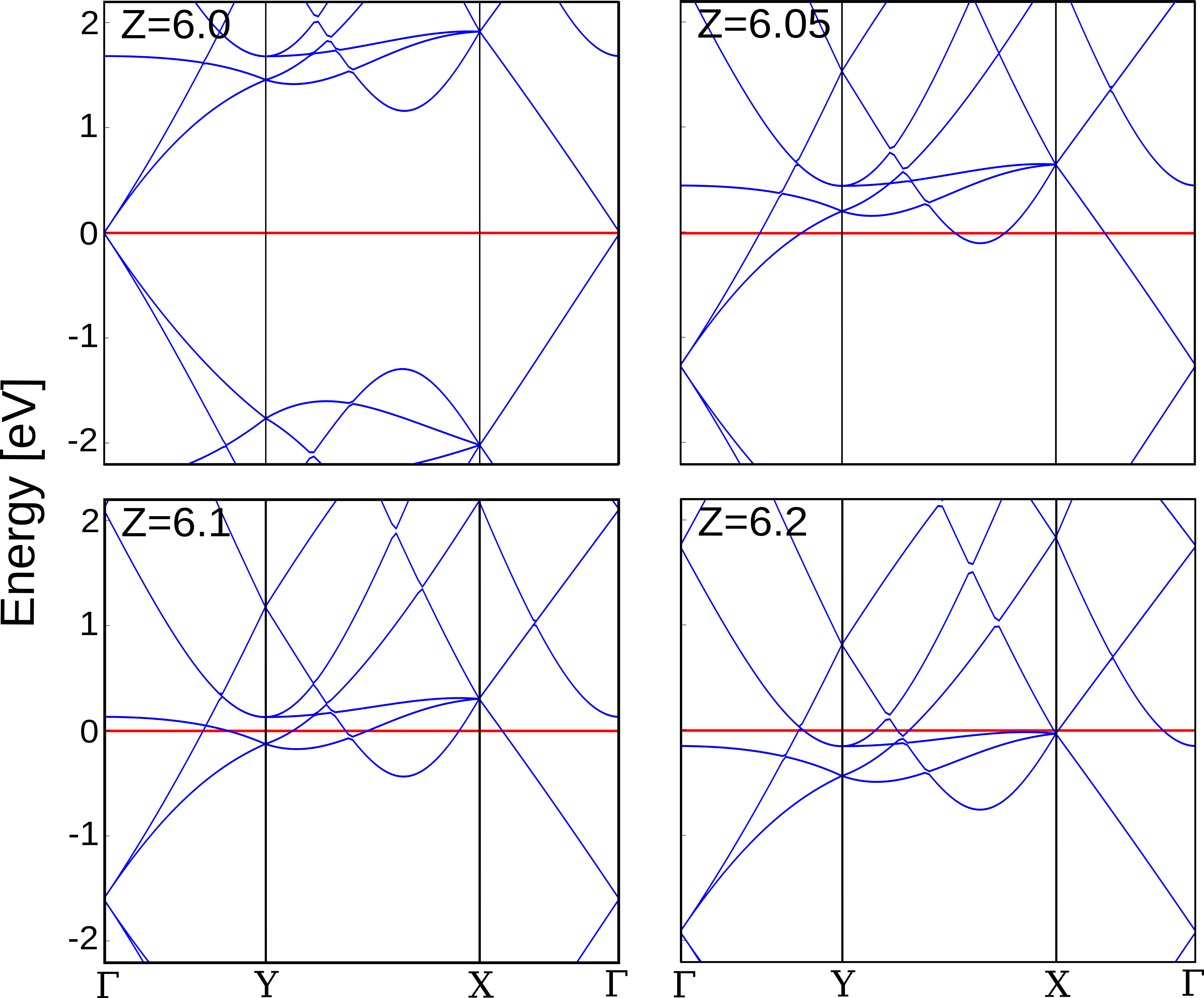}
  \caption{Electronic band structure of the virtual pristine graphene crystal for virtual species with proton/electron number of 6.0$\lesssim Z \lesssim$ 6.2. The zero level (red) refers to the Fermi energy in each system.} 
\label{fig6}
\end{figure}

The second experiment, the virtual crystal approximation, simulates a completely homogeneous addition of electrons, namely at every site in the system. We want to point out that replacing all carbon atoms 
with a virtual species of nuclear charge of 6.1 is not the same as adding 0.1 free electrons per atom, as a certain fraction is localized at the position of the atom due to the increased Coulomb 
attraction of the nucleus. However, a certain fraction is quasi-free and may redistribute according to local variations of the electrostatic potential. The virtual crystal for values 
of 6.0$\lesssim Z \lesssim$ 6.2 retains sp$^2$ hybridized graphene structure. This is reflected in the band structure (Fig.~\ref{fig6}), which remains almost unchanged for values 6.0$\lesssim Z \lesssim$ 6.2
compared to the real carbonic graphene. The effect is mainly a filling of former unoccupied bands together with the corresponding shift of the Fermi level towards n-doped conditions. 
The shift of Fermi energy from 6.0 to 6.05 is larger than from 6.05 to 6.2. This can be explained from the different density of states being present to accommodate the additional electrons. 
Differences in occupation of the bands integrate exactly to the number of additional electrons in the system. Hence, the virtual crystal approximation is a way to add electrons without distorting the electronic structure and therefore proves adequate 
to mimic global doping.
Similar shifts towards p-doped conditions can be achieved analogously with Z$<$6. According to our findings in the previous section, such doping regimes do not promise improved mechanical properties and 
are therefore neglected in the further discussion.

\begin{figure}
\centering
\includegraphics[width=8 cm]{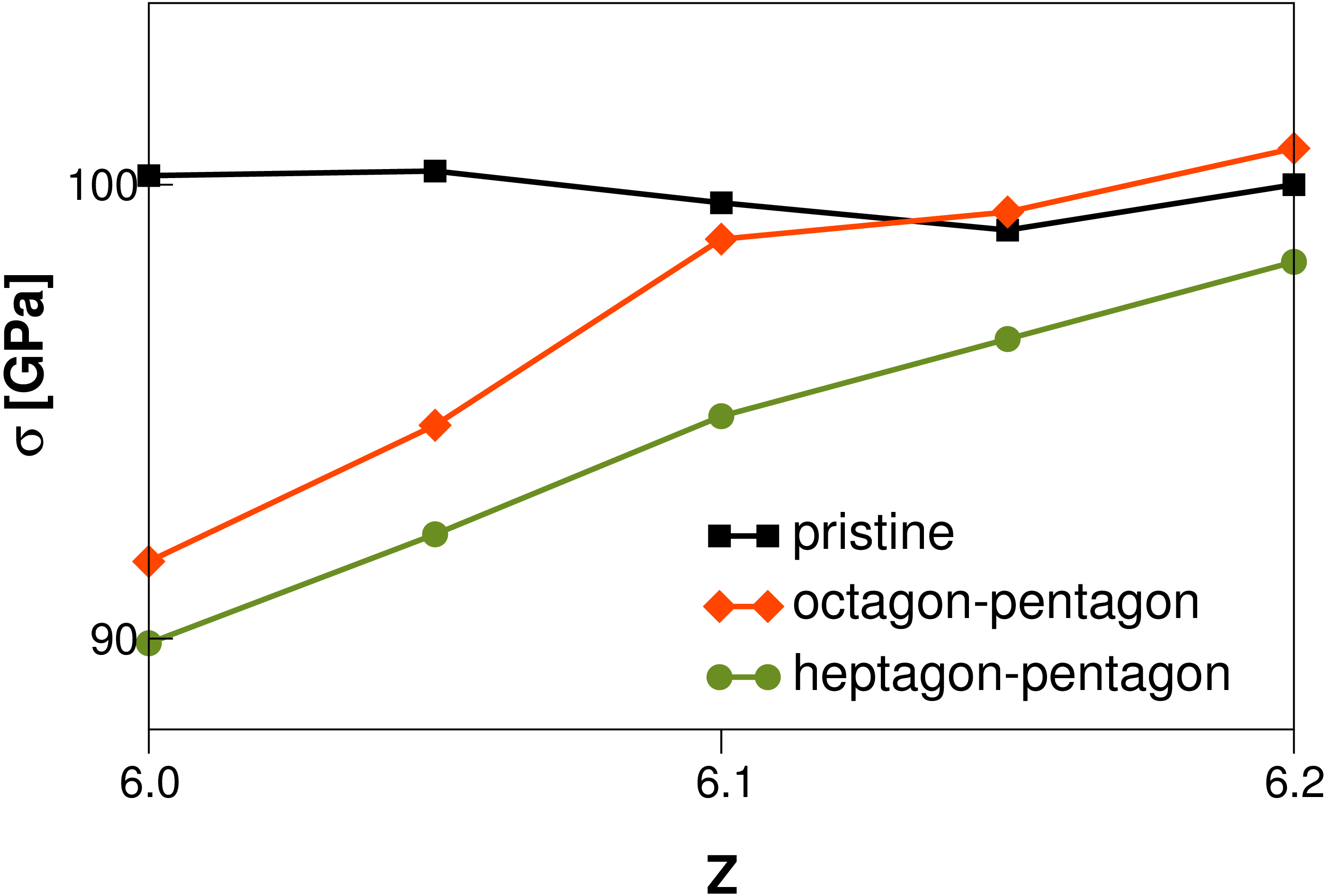}
  \caption{Ultimate yield stress for the octagon-pentagon and heptagon-pentagon line defect and pristine graphene calculated in the virtual crystal approximation. The x-axis labels the proton and electron number of 
  the virtual species constituting the structure. 6.0 refers to the reference real carbon system.} 
\label{fig7}
\end{figure}

The virtual crystal approximation is now applied to study the effect of such additional electrons which arise from global and homogeneous doping on the mechanical properties of line defects.
As can be seen in Fig.~\ref{fig7}, additional electrons only show an effect in case of the line defects, namely an increase of the ultimate strength of about 10\%. 
This can be understood from the frontier molecular orbitals of the line defects which become occupied with increasing Z. As discussed previously in Section~\ref{secIIIA}, these states are located in the defect region
and promote chemical bonding.
The ultimate strength of the pristine structure of virtual species with 6.0$\lesssim Z \lesssim$ 6.2 is largely the same as of the real carbonic system. 
The increase in ultimate strength is therefore solely due to the stabilization of the defect region. We note that for $Z >$ 6.2 the ultimate strength of the pristine structure increases significantly 
(not shown in the Fig.~\ref{fig7}). However, for such large values of Z the band structure shows more significant deviations from the Z=6.0 case. We therefore do not consider this feature a physical effect but rather the break-down 
of the virtual crystal approximation.

For the validated values of 6.0$\lesssim Z \lesssim$ 6.2, the improved mechanical properties have to be regarded as a true physical effect of additional charges localized in the defect region.
This suggests that strengthening of the defected graphene structures can not only be achieved through substitutional doping at specific sites, but also
through a global shift of the Fermi energy. This can alternatively be realized through global doping or application of an external voltage.


\section{Summary and Conclusions}
In summary, we performed an \textit{ab-initio} computational study of the electronic structure and mechanical properties of the octagon-pentagon and heptagon-pentagon line defect in 
graphene. We found that the electronic properties of the system change significantly upon application of external strain. Specifically, we predict 
a strain-induced transition from semi-conducting to metallic behaviour for the heptagon-pentagon line, which promises interesting technological application. 
We scrutinized the effect of substitutional nitrogen (boron) doping with different concentrations and determined sites A,D \& H (only H for boron) as thermodynamical most accessible doping sites.
We found that nitrogen is energetically much more favored than boron as a dopant. 
We showed that substitutional nitrogen doping shifts the Fermi energy while leaving the band structure largely untouched, which suggests doping as a potential pathway for electronic structure engineering, while maintaining
the electronic properties of the pristine graphene domain.
Substitutional nitrogen doping was furthermore found to enhance the ultimate strength, eliminating the line defect as the weak spot in the graphene structure.
We disentangled local and global doping effects by comparing to substitution with virtual species in the virtual crystal approximation.  
We argue that just like n-doping substitutional nitrogen doping provides quasi-free electrons which occupy localized states in the defect region and thus promote the ultimate strength of the material.
Weakening through p-doping can be explained from the intrinsic electron affinity of the line defect. 
Competition of the boron atom with the defect region as the stronger electron sink depopulates these defect states, which compromises the mechanical properties.
We conclude that stabilization of the defect region can also be realized through application of an electric potential, an effect which we suggest to exploit in graphene devices under extreme conditions.


\section{Acknowledgements}
We acknowledge access to the supercomputing facilities of IDRE and MIRA. This work was funded by the National Science Foundation under Grant NSF-CHE 1125931.

%

\end{document}